\documentclass[article,showpacs,showkeys,superscriptaddress]{revtex4}
\usepackage{amsmath}
\usepackage{bbold}
\usepackage{amsfonts}
\usepackage{amssymb}
\usepackage{pbsi}
\usepackage[T1]{fontenc}
\usepackage{xcolor}
\usepackage{graphicx}
\usepackage{subfig}
\usepackage{float}
\usepackage{color}
\usepackage{hyperref}
\hypersetup{colorlinks = true,
	linkcolor = red,
	anchorcolor = blue,
	citecolor = blue,
	filecolor = blue,
	urlcolor = blue}

\begin{document}

\title{Electromagnetic plasma waves in Big Bounce and Big Crunch cosmologies}

\author{Felipe A. Asenjo}
\email{felipe.asenjo@uai.cl}
\affiliation{Facultad de Ingenier\'ia y Ciencias,
Universidad Adolfo Ib\'a\~nez, Santiago 7491169, Chile.}

\date{\today}

\begin{abstract}
We study the exact dynamics of electromagnetic  waves in cold electron-positron plasma in the simplest background of Big Bounce and Big Crunch cosmologies. We show that these waves are described by a Mathieu equation in a Big Bounce cosmology, which opens the possibility of electromagnetic wave amplification due to the cosmological constant in this Universe. On the contrary, in  a Big Crunch cosmology, electromagnetic plasma waves are described by a modified Mathieu equation. In both cases, the features (grow/collapse) of the Universes imprint similar characteristics in the temporal evolution of these waves.

\end{abstract}



\maketitle

\section{Introduction}

Plasma waves have been extensively studied in different cosmological contexts \cite{in1,in2,in3,in4,in5,in6,in7,in8,in9,in10,in11,in12,in13,in14,asenjo2}.  The evolution of the Universe, through the scale factor, modifies the propagation properties of the plasma waves. However, most of those works have been carried out for standard cosmological models. In this work, very interestingly,  we show that electromagnetic plasma waves evolves in an exact manner in a Universe filled only with (non-spatially flat) curvature and cosmological constant, producing
a Big Bounce or a  Big Crunch cosmology.  The evolution of this Universe is described by the Friedman-Robertson-Walker metric, with an isotropic scale factor $a$, and  the equation \cite{ryder}
\begin{equation}
    \frac{H^2}{H_0^2}=\Omega_0+\frac{1-\Omega_0}{a^2}\, ,
    \label{hubb}
\end{equation}
where $H=\dot a/a$ (with $\dot a=da/dt$), $H_0$ is the Hubble constant (at present-day), and $\Omega_0$ is the density parameter measuring the sign of curvature of the Universe. For a model of an Universe with only dark energy and  a non-spatially flat curvature, we have $\Omega_0\neq 1$, and 
$\Omega_0=\Omega_{\Lambda,0}$, where $\Omega_{\Lambda,0}$ is the present-day density parameter of the cosmological constant contribution. 

A Big Bounce cosmology  can be obtained from Eq.~\eqref{hubb} when the Universe has positive curvature, with $\Omega_0>1$. In this case,  the solution of Eq.~\eqref{hubb} is \cite{ryder}
\begin{equation}
    a(t)=a_B \cosh\left[\sqrt{\Omega_0} H_0 (t-t_B)\right]\, ,
    \label{BBa}
\end{equation}
where $a_B=\sqrt{({\Omega_0-1})/{\Omega_0}}<1$, and
$t_B$ is the time when the Big Bounce occurred. For this case, time runs from $0\leq t\leq t_0=t_B+{\mbox{arccosh}}(1/a_B)/(\sqrt{\Omega_0} H_0)$, such that $a(0)=a_B \cosh(\sqrt{\Omega_0} H_0t_B)$, and $a(t_0)=1.$ Thus, this Universe has a minimum  scale factor $a(t_B)=a_B$, at $t=t_B<t_0$.

On the other hand, in a Universe described by Eq.~\eqref{hubb} with 
negative spatial curvature, $\Omega_0<0$, the resulting dynamics is a Big Crunch cosmology. To obtain the solution of this cosmology we need to do the change  $\Omega_0\rightarrow-|\Omega_0|$ in Eq.~\eqref{BBa}, implying
also that $\sqrt{\Omega_0}\rightarrow i\sqrt{|\Omega_0|}$.
The scale factor is then
\begin{equation}
    a(t)=a_C \cos\left[\sqrt{|\Omega_0|}\, H_0 (t-t_C)\right]\, ,
    \label{BCa}
\end{equation}
where $a_C=\sqrt{(|\Omega_0|+1)/|\Omega_0|}$, and $t_C=\pi/(2 \sqrt{|\Omega_0|}H_0)$ as the time when the Big Crunch starts to occur. Thus $a(0)=0$, to later the Universe grow to a maximum value $a_C$ at time $t_C$, to  finally collapse with $a(t_f)=0$, at final time $t_f=2 t_C$. The crunch of the Universe happens at half of its whole evolution time.

In the following section we outline the general theory for electromagnetic plasma waves in any cosmological scenario. In Sec. III we study those waves in a Big Bounce cosmology, to  later study them in a Big Crunch cosmology in Sec. IV. 

\section{Electromagnetic plasma waves in cosmology}

We will model the plasma dynamics in the cosmological background given by scale factors \eqref{BBa} and \eqref{BCa}. In particular, we study electromagnetic plasma wave propagation.
In a cold relativistic plasma, in a general cosmological background,
the electromagnetic plasma waves in an electron-positron plasma 
ca be put in terms of the evolution of the magnetic field ${\bf B}$. This follows the equation \cite{asenjo1,asenjo2}
\begin{equation}
    \frac{1}{a^2}\frac{\partial}{\partial t}\left(a\frac{\partial}{\partial t}\left(a^3 {\bf B}\right)\right)-\nabla^2{\bf B}+\frac{\omega_p^2}{a}{\bf B} =0\, .  
    \label{eqB1}
\end{equation}
where $\omega_p$ is the plasma frequency of the electro-positron system. Eq.~\eqref{eqB1} can be simplified by written it in terms of the conformal time
\begin{equation}
    \tau=\int \frac{dt}{a}\, ,
    \label{conformal}
\end{equation}
and by defining the effective magnetic field ${\bf b}$ through the relation
\begin{equation}
    {\bf B}(\tau,{\bf x})=\frac{{\bf b}(\tau)}{a^3}\exp(i\, {\bf k}\cdot{\bf x})\, ,
    \label{effectivebfield}
\end{equation}
where ${\bf k}$ is a wavenumber vector of the electromagnetic plasma wave. Thus, the temporal evolution of the effective magnetic field is given by
\begin{equation}
    \frac{d^2 {\bf b}}{ d\tau^2}+\left(k^2+\frac{\omega_p^2}{a}\right){\bf b} =0\, ,
    \label{eqB2}
\end{equation}
where $k^2={\bf k}\cdot{\bf k}$, and $a$ must be written in terms of conformal time $\tau$. The solutions of Eq.~\eqref{eqB2} complete the  evolution in time of cosmological electromagnetic plasma waves. 

In the following we show its exact dynamics for Big Bounce and Big Crunch
cosmological scenarios.

\section{Electromagnetic waves in a Big Bounce cosmology}

An Universe of this kind, with positive curvature and metric \eqref{BBa}, describe a Big Bounce cosmology. The conformal time
\eqref{conformal}
is written as
\begin{equation}
   \tau=\frac{1}{\chi}\arctan\left[\sinh\left(\sqrt{\Omega_0} H_0(t-t_B) \right)\right]\, .
\end{equation}
where $\chi= \sqrt{\Omega_0-1}\, H_0$ is the effective conformal frequency. The conformal time runs in the form $\tau_i\leq \tau\leq \tau_f$, where $\tau_i=-\arctan[\sinh(\chi t_B/a_B)]/\chi$, and $\tau_f=\arctan[\sqrt{(1/a_B)^2-1}]/\chi$, and where the bounce occurs in $\tau=0$. 

By using this conformal time, we can re-write the scale factor  \eqref{BBa} as
\begin{equation}
    a(\tau)=\frac{a_B}{\cos\left(\chi\tau\right)}\, .
    \label{scalefactorBB}
\end{equation}
We can use now this scale factor in Eq.~\eqref{eqB2} to describe electromagnetic plasma waves in a Big Bounce cosmology. The equation takes the form
\begin{equation}
    \frac{d^2 {\bf b}}{ dy^2}+\left(\frac{k^2}{\chi^2}+\frac{\omega_p^2}{\chi^2 a_B}\cos y\right){\bf b} =0\, ,
    \label{eqB3}
\end{equation}
where $y=\chi\tau$. 
The dynamics described by Eq.~\eqref{eqB3} is thus given by a Mathieu equation. 
There
is vast literature on the exact different type of solutions of this equation \cite{bender}. 
Therefore, all the temporal behavior of electromagnetic plasma waves in a Big Bounce cosmology
is already exactly known depending on the different cosmological and plasma parameters. For arbitrary combination of those parameters, the solutions are stable and periodic, with some varying amplitude. In particular, exact expressions for stable solutions are known for small values of the parameter ${\omega_p^2}/({\chi^2 a_B})$. Thus, in general, the  electromagnetic plasmas waves are periodic waves in this cosmology. 

\begin{figure}
    \centering    
\includegraphics[width=0.6\linewidth]{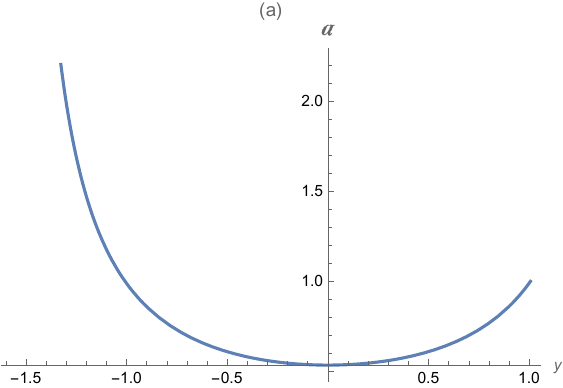}\\ \includegraphics[width=0.6\linewidth]{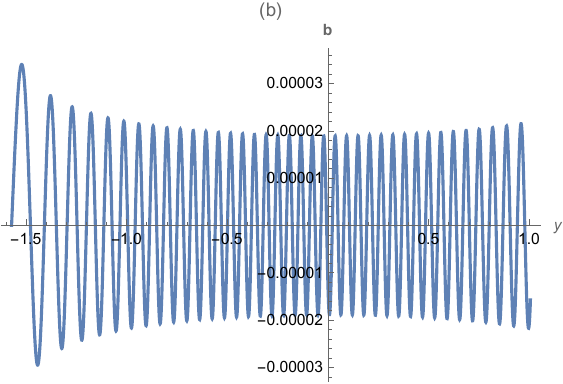}\\    \includegraphics[width=0.6\linewidth]{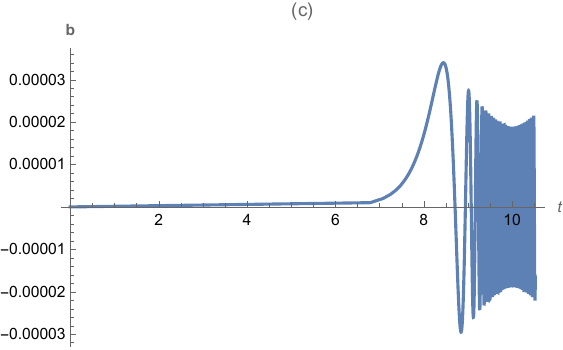}
       \caption{(a) Evolution of scale factor \eqref{scalefactorBB} for $\Omega_0 = 1.4$, $H_0 = 2$, $k/\chi \approx 24$,   ${\omega_p^2}/({\chi^2 a_B})\approx 11690$, $\tau_i\approx -1.571$, and $\tau_f\approx 1.007$.
       (b) Particular periodic solution of Eq.~\eqref{eqB3} for ${\bf b}(0)=0$, $d{\bf b}(0)/dy=0.001$, with same parameter than (a).  (c) Solution for ${\bf b}$ in terms of time $t$, with the same parameters than (a).}
    \label{solfigs}
\end{figure}

An example of a numerical solution of Eq.~\eqref{eqB3} is shown in Fig.~\ref{solfigs}(b), where a periodic solution is obtained for arbitrary values of $\Omega_0 = 1.4$, $H_0 = 2$, $k/\chi \approx 23.7$,  ${\omega_p^2}/({\chi^2 a_B})\approx 11692.7$, and $t_B=10$. This implies that $\tau_i\approx -1.571$, and $\tau_f\approx 1.007$. Compare this solution with the temporal evolution of scale factor \eqref{scalefactorBB} shown in Fig.~\ref{solfigs}(a).
The wave shows a bounce around $y=0$ ($\tau=0$), where the Big Bounce of the Universe takes place. Notice that the temporal dynamics of  ${\bf b}$ follows the evolution of this Big Bounce Universe. As the Universe decreases (increases) its scale factor, the amplitude of 
${\bf b}$ decreases (increases). The behavior of the electromagnetic field ${\bf b}$ in terms of time $t$ is shown in Fig.~\ref{solfigs}(c) for the same values of parameters of Fig.~\ref{solfigs}(b). As the Universe evolves into the Big Bounce, the electromagnetic wave amplitude first grows to later decreases. Then, it experiences the bounce on its amplitude induced by the cosmology.

Despite of the general periodic solutions, there are certain values on the parameter phase space in which a parametric instability (amplification) is triggered in the wave by the Big Bounce cosmology. The first
relevant unstable
region for the electromagnetic wave dynamics 
occurs when $\omega_p^2/(\chi^2{a_B})\ll 1$,  in the vicinity of the values ${k}^2/{\chi}^2= {1}/{4}\pm \omega^2/(2\chi^2 a_B)$ \cite{bender}. This implies that waves with wavelengths $\lambda=2\pi/k$ can be amplified by this cosmology when they have values of the order
\begin{equation}
    \lambda= \frac{4\pi}{\chi}\left(1\pm \frac{\omega_p^2}{\chi^2 a_B}\right)\, .
    \end{equation}
Under this specific value of $k$ (or $\lambda$), the electromagnetic plasma wave experiences an instability, being therefore  amplified  exponentially (in times $\tau$) at expenses of the evolution of the cosmology, in the form  \cite{bender} 
\begin{equation}
    |{\bf b|}\propto \exp\left(\frac{\omega_p^2}{2\chi^2 a_B}y\right)=\exp\left(\frac{\omega_p^2}{2\chi a_B}\tau\right)\, .
\end{equation}
Therefore, this amplification can be achieved for a very diluted plasma, in a Universe with highly space curvature, $\Omega_0\gg 1$. It is essential the interaction of the plasma (through $\omega_p$) with this Big Bounce cosmology to trigger the amplification of the electromagnetic wave.


\section{Electromagnetic waves in a Big Crunch cosmology}

For a Big Crunch cosmology with scale factor \eqref{BCa}, the conformal time \eqref{conformal} becomes
\begin{equation}
   \tau=\frac{1}{\xi}\mbox{arctanh}\left[\sin\left(\sqrt{|\Omega_0|} H_0(t-t_C) \right)\right]\, .
\end{equation}
where $\xi= \sqrt{|\Omega_0|+1}\, H_0$ is the corresponding conformal frequency of this cosmology. In terms of the conformal time, the evolution of this Universe goes  $-\infty\leq \tau\leq\infty$, where the maximum value for the scale factor is reached at $\tau=0$.

In this way, the scale factor \eqref{BCa} becomes written in terms of the conformal time as
\begin{equation}
    a=a_C\,  \mbox{sech}\left(\xi \tau\right)\, .
    \label{scalefactorBC}
\end{equation}
Using this scale factor in Eq.~\eqref{eqB2}, we can now describe the temporal evolution of cold electromagnetic plasma waves in a Big Crunch cosmology, as
\begin{equation}
    \frac{d^2 {\bf b}}{ dy^2}+\left(\frac{k^2}{\xi^2}+\frac{\omega_p^2}{\xi^2 a_C}\cosh y\right){\bf b} =0\, ,
    \label{eqB4}
\end{equation}
where now $y=\xi\tau$. 
In this cosmology, the evolution of the electromagnetic plasma waves is described by a modified Mathieu equation \cite{ruby}.
This equation can be solved numerically, and the behavior of their solutions is anew  known. However, there are not periodic solutions of this equation \cite{ruby}.

\begin{figure}
    \centering    
\includegraphics[width=0.6\linewidth]{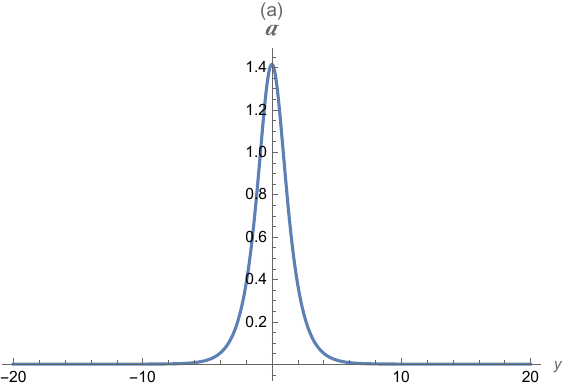}\\\includegraphics[width=0.6\linewidth]{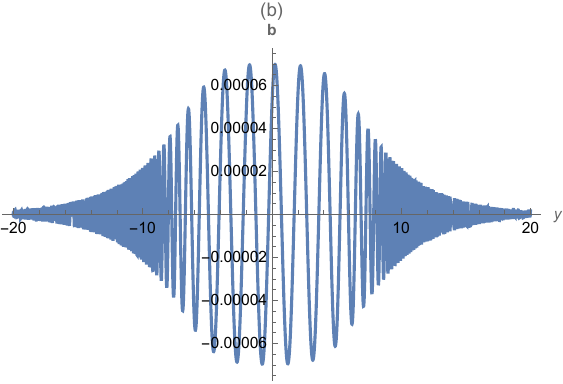}\\
\includegraphics[width=0.6\linewidth]{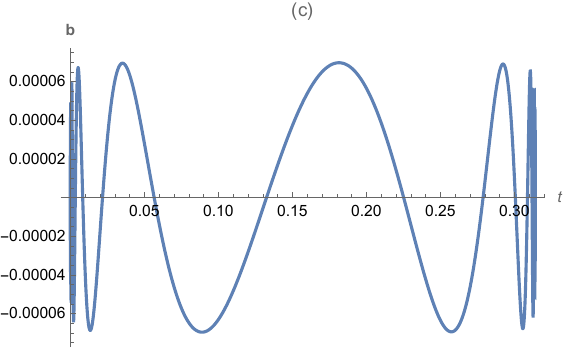}
       \caption{ (a) Evolution of scale factor \eqref{scalefactorBC}, for $|\Omega_0| = 1$, $H_0 = 10$, $k/\chi \approx \sqrt{10}$,   and ${\omega_p^2}/({\xi^2 a_C})\approx \sqrt{0.1}$. (b) Particular  solution of Eq.~\eqref{eqB4},  for ${\bf b}(0)=0$, $d{\bf b}(0)/dy=0.01$, and same parameters than (a). (c) Solution for ${\bf b}$ in terms of time $t$, with the same parameters than (a).}
    \label{solfigs2}
\end{figure}

 In Fig.~\ref{solfigs2}(b), we plot a numerical solution of Eq.~\eqref{eqB4}, in terms of time $y$, for ${\bf b}(0)=0$, and $d{\bf b}(0)/dy=0.01$, with the arbitrary parameters $|\Omega_0| = 1$, $H_0 = 10$, $k/\chi \approx \sqrt{10}$,   and ${\omega_p^2}/({\xi^2 a_C})\approx \sqrt{0.1}$. These values imply a crunch time $t_C=\pi/20$, and a total evolution time $t_f=\pi/10$. The electromagnetic wave is amplified till the crunch time to later decrease as the Universe decreases. Compare this behavior with the temporal evolution of the Universe depicted in Fig.~\ref{solfigs2}(a).
Also, in Fig.~\ref{solfigs2}(c),
we plot the same solution for the evolution of the electromagnetic wave in terms of time $t$. The main effect occurs in the frequency of the wave, which decreases till the crunch time, to later increase as the plasma wave is compressed by the Universe.

\section{Final remark}

With the previous calculations we have shown that electromagnetic plasma waves have exact dynamics in each of these cosmologies. The family of their solutions are the ones of Mathieu equations.
In general, the main features of the the plasma waves follow the temporal evolution of the cosmologies. In both cases, as the Universe scale factor decreases (increases), the amplitude of the electromagnetic wave also decreases (increases). An inverse behavior takes place for the frequency of the plasma wave; it increases (decreases) as the scale factor decreases (increases). 

A more detailed analysis of Eqs.~\eqref{eqB3} and \eqref{eqB4} is left for the future, in where the complete family of solutions, for the whole space of plasma parameters, will be studied.

\begin{acknowledgements}
FAA thanks to FONDECYT grant No. 1230094 that partially supported this work.
 \end{acknowledgements}

\end{document}